\documentclass[12pt, eqsecnum,preprint]{aastex}
\slugcomment{August 21, 2002}

\usepackage{amsmath}
\usepackage{subeqn}
\usepackage{graphicx}
\usepackage{subfigure}
\usepackage{tensor}

\def\ahalf{{\frac{1}{2}}}

\def\prn#1{{\left(#1\right)}}
\def\brk#1{{\left[#1\right]}}
\def\brc#1{{\left\{#1\right\}}}

\begin{document}

\title{RELATIVISTIC SINGULAR ISOTHERMAL TOROIDS}

\author{Mike J. Cai}
\affil{Department of Physics, University of California \\
              Berkeley, CA 94720-3411, USA}
\email{mcai@astron.berkeley.edu}
\author{Frank H. Shu}
\affil{Physics Department, National Tsing Hua University\\
              Hsinchu 30013, Taiwan, ROC}
\email{fshu@astron.berkeley.edu}

\begin{abstract}
We construct self-similar, axisymmetric, time-independent
solutions to Einstein's field
equations for an isothermal gas with a flat rotation
curve in the equatorial plane.  The metric
scales as $ds^2 \rightarrow \alpha^2 ds^2$ under the transformation
$r\rightarrow \alpha r$ and $t \rightarrow \alpha^{1-n} t$, where $n$
is a dimensionless measure of the strength of the gravitational
field.  The solution space forms a two-parameter family characterized
by the ratios of the isothermal sound
speed and the equatorial rotation speed to the speed of light.
The isodensity surfaces are toroids, empty of matter along the rotation axis.
Unlike the Newtonian case, the
velocity field is not constant on a cylindrical radius because of
frame dragging.  As the
configuration rotates faster, an ergoregion develops in the form of
the exterior of a cone centered about the rotation axis.  The sequence
of solutions terminates when frame dragging becomes infinite and the
ergocone closes onto the axis.  The fluid velocity of the last
solution has a modest value in the midplane but reaches the speed of light
on the axis.
\end{abstract}

\newcommand\R[2]{R_{(#1)(#2)}}
\newcommand\T[2]{T_{(#1)(#2)}}
\newcommand\tet[2]{\tensor{e}{_{(#1)}^#2}}

\section{Introduction}

In remarkable treatments of the axisymmetric equilibria of
self-gravitating, isothermal, unbounded, stellar and gaseous
systems with flat rotation curves,
Toomre (1982) and Hayashi et al. (1982)
found completely analytical solutions for these self-similar configurations.
The theory applies in the non-relativistic limit when mechanics and
gravitation can be approached by Newtonian concepts.
Written in spherical polar coordinates $(r,\theta,\phi)$,
the density profile of a Hayashi-Toomre model has the form:
\begin{equation}
    \rho(r,\theta) =
        \prn{1+ \frac{V^2}{2a^2}}^2\prn{\frac{a^2}{2\pi Gr^2}}
            \csc^2\theta \,{\rm{sech}}^2\brk{
            \left(1+\frac{V^2}{2a^2}\right)
            \ln \cot \frac{\theta}{2}},\label{T-H}
\end{equation}
where $V$ is the rotation velocity and
$a$ is the isothermal acoustic (or stellar dispersive) speed.
Notice that, apart from a trivial scaling relative to $a^2$,
the Hayashi-Toomre models form a {\it linear sequence} characterized
by the single value $V^2/a^2$.  Notice also that $\rho=0$ for
$\theta=0$ or $\pi$, whereas $\rho \rightarrow \infty$ as $r\rightarrow 0$
and the enclosed mass within a sphere of radius $r$ goes to infinity
as $r\rightarrow \infty$.  These properties account for our
assignment of the name ``singular isothermal toroid'' (SIT) for
this generic class of models (see also Li \& Shu 1996).

What is the relativistic generalization
of the Hayashi-Toomre sequence when $a$ and $V$ are not
small compared to the speed of light $c$?  In this paper
we confirm the intuitive expectation that the linear sequence
will broaden into a two-dimensional surface characterized by
the pair of dimensionless parameters $v\equiv V/c$ and
$\gamma \equiv a^2/c^2$.  In the limit $v^2/\gamma \equiv V^2/a^2 \gg 1$,
we further anticipate that relativistic SITs will become
highly flattened, like their non-relativistic analogs in the same limit;
i.e., SITs will become SIDs (singular isothermal disks).\footnote{In
this paper, we
use the terminology ``isothermal'' somewhat loosely to mean
that the pressure is directly proportional to mass density $\rho$,
with a constant of proportionality that we call $a^2 \equiv \gamma c^2$.
Such a proportionality can arise, of course, in physical systems in a wider
context than when the thermodynamic temperature is a strict constant.}

Relativistic SIDs have been studied by
Cai \& Shu (2002, hereafter CS), who took
their inspiration from the cold-disk work of Lynden-Bell \&
Pineault (1978a,b).  CS adopted the simplification
of negligible disk thickness by assuming an anisotropic
pressure tensor, with zero and nonzero effective sound speeds in the
vertical and horizontal directions respectively.
Part of the rationale for the present
investigation is to ascertain the range
of validity of the disk
approximation when the same effective sound speed $a$ applies
in all three principal-axis directions, and when the
combination $v^2/\gamma$ is not necessarily large compared to unity.
In particular, we wish to examine how frame-dragging
may distort expectations of toroid flatness
formed by naive extrapolations from the Newtonian experience.

The astronomical motivation for the work of Hayashi et al. (1982)
and Toomre (1982) comes from interstellar gas clouds and
disk galaxies.  It is unknown whether such objects have their
relativistic counterparts in astrophysics.  For example,
active galactic nuclei and quasi-stellar objects are increasingly believed
to be powered by supermassive blackholes (SMBHs).  Are such SMBHs
the central products of slow growth through an accretion disk?
Or do they result from successive mergers of smaller units
(stars or smaller BHs) in the nuclei of galaxies?
Or could SMBHs have formed from the monolithic gravitational collapse
of relativistically compact gaseous or stellar systems that
resemble SITs?

Haehnelt \& Kauffmann (2001) discuss the difficulties associated
with the first two scenarios.  If the last scenario is a viable or
even likely possibility, it would be important to develop
theoretical predictions and observational tests.
Such tests provide another motivation for the present line
of research.  In particular, it is known that slowly rotating,
non-relativistic, gaseous SITs are unstable to inside-out
gravitational collapse to form a steadily growing pointlike mass
at the center (see, e.g., Shu 1977 for the simplest case of the
collapse of a singular isothermal sphere or SIS).  The analogous
collapse of a suitably flattened, relativistic SIT (during the
earliest epochs of galaxy formation) to form a steadily growing
SMBH (with a Kerr-like geometry) should be accompanied by the
copious generation of gravitational radiation.  Such radiation
might be detectable by future gravitational-wave observatories.

Differentially rotating objects exemplified by SITs and SIDs have
other advantages over the uniformly rotating, relativistic disks
investigated in the pioneering work of Bardeen \& Wagoner (1971).
From numerous studies of self-gravitating systems in the Newtonian
regime (see, e.g., Toomre 1977, Binney \& Tremaine 1978, Lowe et
al. 1994, Bertin et al. 1989ab, Goodman \& Evans 1999, Shu et al.
2000), it is known that the transfer of angular momentum from
inside to outside through spiral and barlike density waves is a
natural phenomenon in configurations that are differentially
rather than uniformly rotating.  Moreover, for similar
distributions of specific angular momentum, differentially
rotating systems have lower specific energy, making them more
likely states to be encountered in many natural settings (Mestel
1963). These considerations add yet more motivation for the
continued theoretical investigation of relativistic SITs and SIDs.

The rest of this paper is organized as follows.  In \S
\ref{geometry}, we develop the mathematical equations governing
SITs.  In \S \ref{Newton}, we solve the Einstein equations
in the Newtonian limit and show that the solution thus
obtained is identical to those of Toomre (1982) and Hayashi et al.
(1982).  In \S \ref{relSIS} we derive analytical solutions for
relativistic SISs when spherical symmetry holds.
\S \ref{3d-Numerical} describes the numerical strategy
we use to solve the fully nonlinear Einstein equations
in the presence of finite rotation.  \S
\ref{3d-Results} summarizes the results of our calculations, including
the three dimensional
velocity field and the degree of flattening due to rotation.
Finally, we offer conclusions and commentary in \S \ref{conclusion}.

\section{Basic Equations}\label{geometry}

\subsection{Metric}

Unlike the razor-thin disks studied in CS, we wish to
consider pressure tensors in this paper that are isotropic
in three dimensions.  In spacetime, the stress-energy tensor
is then a smooth function of polar angle $\theta$,
being positive-definite from pole to equator,
with reflection symmetry about the latter.
Because the sources of gravity are smoother than
a delta function, the functions, $N(\theta)$, $P(\theta)$,
$Q(\theta)$, and $Z(\theta)$ in the
the metric adopted by CS,
\begin{equation}
    ds^2 = -r^{2n} e^N dt^2 + r^{2} e^{2P-N} (d\phi - r^{n-1} e^{N-P}
Q dt)^2 + e^{Z-N} (dr^2 + r^2 d\theta^2),\label{metric}
\end{equation}
have no kinks as we cross the midplane, unlike the disk case. For
clarity, we recapitulate a few important properties of this
metric.

In the above equation $n$ is a number between 0 and 1, which we
call the gravitational index.  In our choice of coordinates (but
not in Lynden-Bell \& Pineault 1978a,b), the radial coordinate $r$
has the usual dimensions of length.  The presence of the extra
factor $r^n$ in front of each power of $dt$ would then seem to
indicate that $t$ does not have the unit of length (in geometric
units where $c$ = $G$ = 1).  This impression is mistaken. As
explained in CS, the gravitational potential of self-similar SISs,
SIDs, or SITs have logarithmic dependences on the radial
coordinate $r$. For example, except for slight notational
modifications, Toomre (1982) and Hayashi et al. (1982) found that
the gravitational potential associated with the density
distribution (\ref{T-H}) is given in conventional units by
\begin{equation*}
    \Phi = (V^2+2a^2)\ln r + \ahalf c^2 N(\theta),
\end{equation*}
where $N(\theta)$, up to an additive constant, has a functional form
that can be obtained by integration of the expression for $N'$ to
be given in \S \ref{Newton}.  When exponentiated,
as in $g_{tt} = -e^{2\Phi/c^2} = -r^{2n}e^{N}$ with $n=v^2+2\gamma$,
this logarithmic dependence yields one troublesome factor of $r^n$
for each index of $t$ involved in a metric coefficient
associated with the differential of time.
However, an arbitrary length scale $L$ introduced to make the
argument of the logarithmic term formally dimensionless, i.e., to
convert $\ln r$ to $\ln (r/L)$ would add an arbitrary constant
to the potential $\Phi$ that would have no physical conseqences.
The factors of $r^n$ or $r^{2n}$ then become the dimensionless
combinations $(r/L)^n$ or $(r/L)^{2n}$, which would restore to
time $t$ its usual unit.  The arbitrariness of $L$ then
merely reflects the freedom to scale in a problem that
is self-similar and lacks intrinsic length scales.
For notational compactness, we have adopted the convention of
setting $L$ equal to unity.  In any contour-level figure, therefore,
the reader can make whatever
length assignment she or he might like to any given contour,
and scale all other
contours accordingly.  (Except for size, they all look the same.)

The coordinate $\theta$ differs from the normal
definition of colatitude by a constant factor $k\ge 1$
(see again the discussion of CS).  If $\theta$
were defined as usual to have a range from 0 to $\pi$
(as in Lynden-Bell \& Pineault 1978a,b), $k$ would
appear elsewhere in the problem as an eigenvalue to be
found by satisfying certain boundary conditions.
Such eigenvalue searches in
nonlinear differential equations are numerically very expensive.
By absorbing the factor $k$ into the definition of $\theta$, we
change the nature of the eigenvalue search to placing the midplane
in a proper location $\theta_0\neq \pi/2$.
Determining the value of $\theta_0$ turns out to be possible via
a computationally non-taxing shooting scheme in our formulation of
the numerical solution of the Einstein equations.

Physically, the point is as follows.  The gravitation of a
flattened distribution of matter curves space in such a way as to warp
the polar angle $\theta$ if we choose $e^{[Z(\theta)-N(\theta)]/2}r$
to be the radial distance
from the origin along any path of constant $t$, $\phi$ and $\theta$.
(The assumption of axial symmetry allows us to preserve
the usual meaning of the azimuthal angle $\phi$.)  This warping is such
as to make the total angle coverage from north rotational pole to
south rotational pole greater than $\pi$ (but always less than $2\pi$).

Apart from the transformations described above, the metric (\ref{metric})
has the only form consistent with the requirements of self-similarity,
i.e., it satisfies
the scaling law $ds^2 \rightarrow \alpha^2 ds^2$ when $r
\rightarrow \alpha r$, and $t\rightarrow \alpha^{1-n} t$ for a constant $\alpha$.
As explained in CS, the scaling of $t$ arises from the gravitational
redshift associated with climbing out of the potential of this problem.
Limiting the effect to redshifts then imposes the allowable range $0\le n \le 1$,
as we noted earlier and as will be confirmed in detail later.

\subsection{Einstein's Equations}

We denote ordinary differentiation by $\theta$ with a
subscript in this variable, and define ${\cal R}_{(a)(b)}
\equiv r^2 e^{Z-N} R_{(a)(b)}$, where $R_{(a)(b)}$
is the usual Ricci tensor.
We adopt as a convention that numerical indices or indices in
parentheses are tetrad indices, which are raised and lowered with the
Minkowski metric, while Greek indices are vector indices and are
raised and lowered with the metric coefficients of
equation (\ref{metric}).  In this notation, the scaled
Ricci tensor resulting from
the metric (\ref{metric}) has the
nontrivial components:
\begin{equation}
\begin{split}
    &2{\cal R}_{(0)(0)}=N_{\theta,\theta}+N_\theta P_\theta+2n(1+n)-
Q^2\brc{ \brk{ (\ln Q)_\theta-P_\theta+ N_\theta }^2+(1-n)^2},\\
    &2{\cal R}_{(0)(1)} = Q_{\theta,\theta} + Q_\theta P_\theta -
Q\brk{P_{\theta,\theta}-N_{\theta,\theta}
    +(P_\theta-N_\theta)^2 + P_\theta(P_\theta-N_\theta) + 2(1-n)},\\
    &{\cal R}_{(0)(0)} - {\cal R}_{(1)(1)} =
P_{\theta,\theta}+{P_\theta}^2 +(n+1)^2,\\
    &2{\cal R}_{(2)(2)}= N_{\theta,\theta} -Z_{\theta,\theta}+
P_\theta(N_\theta-Z_\theta)+2n(1-n)+Q^2 (1-n)^2,\\
    &2{\cal R}_{(2)(3)} = (n+1) Z_\theta -2n N_\theta + Q^2 (1-n)
\brk{P_\theta- \prn{\ln Q}_\theta - N_\theta},\\
    &2\left[ {\cal R}_{(3)(3)}+{\cal R}_{(0)(0)} -{\cal R}_{(1)(1)}-
{\cal R}_{(2)(2)}\right] =
    2P_\theta Z_\theta - {N_\theta}^2 + 4n^2 +Q^2
    \brc{\brk{N_\theta + \prn{\ln Q}_\theta -
    P_\theta}^2-(n-1)^2},
\end{split}
\end{equation}
in the locally nonrotating observer (LNRO) frame defined as usual by
\begin{equation}
\begin{split}
    \tet{0}{\mu} &= (r^{-n} e^{-\ahalf N}, r^{-1}Q e^{\ahalf
N-P}, 0, 0),\\
    \tet{1}{\mu} &= (0, r^{-1}e^{\ahalf N-P}, 0, 0),\\
    \tet{2}{\mu} &= (0, 0, e^{\ahalf(N-Z)}, 0),\\
    \tet{3}{\mu} &= (0, 0, 0, r^{-1}e^{\ahalf(N-Z)}).
\end{split}
\end{equation}

For the matter part, we adopt an isotropic
pressure.  The pressure $p$ and energy density $\varepsilon$ are
related by
\begin{equation}
    p = \gamma \varepsilon, \label{EOS}
\end{equation}
where $\sqrt{\gamma}$ is the sound speed.  (In conventional units,
$\varepsilon = \rho c^2$, and $a^2=\gamma c^2$.)  In an equilibrium
solution, there is no vertical or radial velocity, and we may
write in the LNRO frame
\begin{equation}
    u^{(a)} = \prn{\frac{1}{\sqrt{1-v^2}}, \frac{v}{\sqrt{1-v^2}},
    0, 0}.
\end{equation}
Unlike the Newtonian case, where the Poincar\'e-Wavr\'e theorem
ensures that the velocity field is independent of $z$ in
cylindrical coordinates (e.g., Tassoul 1978),
we will have to assume here the velocity
$v$ is a function of the polar angle $\theta$.  For a perfect
fluid, the stress energy tensor is given by
\begin{equation}
    T_{(a)(b)} =(\varepsilon + p) u_{(a)} u_{(b)} + p \eta_{(a)(b)}
\end{equation}
Explicitly, the non-zero components are
\begin{equation}
\begin{split}
    &\T{0}{0} = \varepsilon\frac{1 + \gamma v^2}{1-v^2}, \quad \T{0}{1}
    = -\varepsilon\frac{(1+\gamma)v}{1-v^2},\\
    &\T{1}{1} = \varepsilon\frac{\gamma + v^2}{1-v^2}, \quad \T{2}{2} =
    \gamma \varepsilon = \T{3}{3}.
\end{split}
\end{equation}
The equation of motion $\tensor{T}{^{(a)(b)}_{|(b)}}=0$ has two
nontrivial components:
\begin{equation}
\begin{split}
    (\hat r)&:\quad -n+ v^2
    +2\gamma \frac{1-v^2}{1+\gamma}+Qv(1-n)=0,\\
    (\hat \theta)&:\quad v(Q+v) P_\theta- v Q_\theta
    -(\ln \varepsilon)_\theta \gamma\frac{1-v^2}{1+\gamma}
    -\ahalf N_\theta (1+v^2+2Qv)=0.
\end{split}\label{eom}
\end{equation}
The first equation yields radial force balance:
with $-n$ representing gravity; $v^2$ representing
centripetal acceleration (a factor $1/r$ has been cancelled
from all terms); $2\gamma(1-v^2)/(1+\gamma)$ representing the pressure
gradient (the factor 2 comes from differentiating a pressure
that is proportional to $r^{-2}$ while the factor $(1-v^2)/(1+\gamma)$
comes from making various inertial corrections); and
$Qv(1-n)$ representing the effect of the dragging of inertial frames.
The second equation describes a similar balance
in the $\theta$ direction.  In the problem of an infinitesimally thin disk, both
equations are evaluated on the equatorial plane only.  If we
then impose symmetry about the equator, the second equation is
identically satisfied upon integrating across the mid-plane.  For
the 3-D configuration we are considering here, the second equation
provides a non-trivial consistency relationship for the energy density.

\newcommand{\et}{\hat \varepsilon}
To proceed further, let us define
\begin{equation}
    \et = 8\pi r^2\frac{\varepsilon}{(1+n)^2} e^{Z-N}, \quad \Theta =
    (1+n)\theta,
\end{equation}
and denote differentiation with respect to $\Theta$ by primes.
The Einstein equations may now be written as
\begin{equation}
    R_{(a)(b)} = 8\pi \brk{T_{(a)(b)} - \ahalf \eta_{(a)(b)} T}.
\end{equation}
Written out in full, the components are
\begin{align*}
    &\et \frac{1-\gamma v^2 + 3\gamma + v^2}{1-v^2}= N'' +N'
    P'+\frac{2n}{1+n}- Q^2\brc{\brk{\prn{\ln Q}'-P'+ N'}^2
    + \prn{\frac{1-n}{1+n}}^2},\\
    &-2\et v \frac{1+\gamma}{1-v^2} = Q'' + Q'P'
    - Q\brk{P''-N''+(P'-N')^2 + P'(P'-N') +
    \frac{2(1-n)}{(1+n)^2}},\\
    &2\gamma\et = P'' + {P'}^2 + 1,\\
    &(1-\gamma) \et = N'' -Z''+ P'(N'-Z')+\frac{2n(1-n)}{(1+n)^2}
    +Q^2 \prn{\frac{1-n}{1+n}}^2,\\
    &0=Z' -\frac{2n}{1+n} N' + Q^2 \frac{1-n}{1+n}
    \brk{P'- \prn{\ln Q}' - N'},\\
    &4\gamma\et =2P' Z' - {N'}^2 + \prn{\frac{2n}{1+n}}^2 +Q^2
    \brc{\brk{N' + \prn{\ln Q}' -
    P'}^2-\prn{\frac{1-n}{1+n}}^2}.
\end{align*}

With some algebra, we may separate out a part of
Einstein's equations and the equation of motion \eqref{eom} to
put them into a set of ``dynamical'' equations involving second-
and first-order derivatives, respectively, of metric and matter
variables:
\begin{subequations}
\begin{align}
    N'' &= \et \frac{1-\gamma v^2 +3\gamma + v^2}{1-v^2} -N'
    P'-\frac{2n}{1+n}+ F^2
    + Q^2\prn{\frac{1-n}{1+n}}^2,\label{N'}\\
    Q'' &= -\et (2v+Q+Qv^2) \frac{1+\gamma}{1-v^2} - Q'P'
    + Q\brk{
    (1-Q^2)\prn{\frac{1-n}{1+n}}^2 + (P'-N')^2
    - F^2},\label{Q}\\
    P'' &= 2\gamma\et -1 -{P'}^2,\label{P}\\
    (\ln\et)'&= \brc{v^2 P'- v F-\ahalf  N' (1+v^2)}\frac{1+\gamma}
    {\gamma(1-v^2)}- \frac{1-n}{1+n} \brc{N' - QF},\label{log e}
\end{align}\label{dynamic}
\end{subequations}
where $F = QN'+Q'-QP'$.  This set of ordinary
differential equations (ODEs) is supplemented by
a set of constraint equations from the rest of
Einstein equations and equations of motion involving only 
first- and zeroth-order derivatives,
respectively, of metric and matter variables:
\begin{subequations}
\begin{align}
    0 &=-n+v^2+2\gamma\frac{1-v^2}{1+\gamma}+Qv(1-n),\label{force_balance}\\
    4\gamma\et &=\prn{\frac{2n}{1+n}}^2 +P'N'\frac{4n}{1+n} +
    F^2 +2FP'Q\frac{1-n}{1+n} -
    {N'}^2-Q^2\prn{\frac{1-n}{1+n}}^2.\label{energy}
\end{align}\label{constraint}
\end{subequations}

Notice that $Z$ decouples from the other three metric
coefficients.  Its dynamical ODE can be replaced
by one of the equations of motion associated
with the contracted Bianchi identity.  The function $Z$ can be easily
obtained through integration once the other three metric
coefficients $N$, $P$, and $Q$  are found.

Equation \eqref{force_balance} for radial force balance
has the following important implication:
$v$ cannot be a constant,
as in the nonrelativistic case, but must vary with $\Theta$
if the coefficient governing frame dragging $Q$ does.
Moreover, because $Q(\Theta)$ must vanish on the rotational poles,
$\Theta = 0$ and $2\Theta_0$,
$v$ will generally {\it not} be zero at the rotation axis
for general $n$ and $\gamma$ (e.g., $V\neq 0$ for general Hayashi-Toomre SITs).
Indeed, because $(1-n)$ and $Q$ are both positive, $v$ will
achieve its greatest value on the rotational pole, where frame dragging
cannot help centrifugal forces and pressure gradients
to balance self-similar gravity.

As a final comment, notice that although equation
\eqref{energy} can be used to compute energy density $\et$ on
computational grid points, it does not manifestly require
a positive definite value for $\et$.  Fortunately, we know
that the other non-trivial contracted Bianchi identity guarantees
that equation (\ref{energy}) is a redundant relationship,
given the remaining independent equations of the governing set.
Thus, we can use equaton \eqref{log e} to integrate for $\ln \et$
(always obtaining a positive value $\et$ no matter what the sign
of $\ln \et$), and then check numerically that equation \eqref{energy} 
yields a consistent result (which it always does).  Thus, the remaining
equations \eqref{dynamic} and
\eqref{force_balance} form a complete set
to determine the spacetime geometry.

\subsection{Boundary Conditions}

The equations \eqref{dynamic} form a set of first-order non-linear
ODEs in the five variables $\{N', Q, Q', P', \ln \et\}$.  Thus,
five boundary conditions will specify a solution.
As explained below, we have three
boundary conditions each that we desire to impose
at the pole and the equator, and therefore we seemingly have
an overdetermined situation.  In fact, because the location
of the equator is not given a priori, we have a well-posed problem.

Let us start with the rotation axis. As usual, one
wants a circle in the $\phi$ direction
specified by constant $\Theta$ to have vanishingly
small circumference if $\Theta \rightarrow 0$. Thus, we need
$e^P \rightarrow 0$, which requires $P$ to go to $-\infty$
at the pole $\Theta = 0$.  In order to have a non-singular
geometry on the axis, the Ricci tensor must remain finite as
$\Theta \rightarrow 0$.  The boundary conditions on the axis that
are compatible with the Newtonian limit are therefore
\begin{equation}
\begin{split}
    N'=0&:\qquad \text{The gravitational field is smooth across the axis.}\\
    Q=0&:\qquad \text{There is no frame dragging on the axis.}\\
    P'=+\infty&:\qquad \text{Coordinate singularity.}
\end{split}\label{3d-boundary:axis}
\end{equation}

We noted earlier that the fluid velocity $v$ is generally
non-vanishing on the rotation axis (as is true, e.g., in the
Hayashi-Toomre models).  We anticipate therefore that the infinite
``centrifugal'' effect at the pole will drive away all matter from
it and create a cavity there (explaining, e.g., why the
Hayashi-Toomre configurations are ``toroids''). If we suppose that
a near-vacuum situation applies also to relativistic SITs and
approximate $\et$ to be vanishingly small near the rotation axis,
we can show that equation \eqref{log e} has the general solution,
\begin{equation}
P = \ln [K\sin (\Theta-\Theta_1)],
\end{equation}
in a small neighborhood of $\Theta=\Theta_1$ where $\et$ vanishes,
with $K$ and $\Theta_1$ being integration constants.
The coordinate singularity imposed by $P'$ being infinite
on the north pole then identifies that $\Theta_1 = 0$; i.e.,
$P' \approx 1/\Theta$ for small $\Theta$.
Notice finally that the divergence of $P'$ at the north pole is consistent
with the vanishing there
of the energy density $\et$ on the left-hand side of equation (\ref{log e}).
Thus, from a physical point of view, we can now see that
the boundary conditions $Q=0$ and $P'=+\infty$ at the axis are not
two distinct requirements, but only one (once we fix the
north pole to be at $\Theta=0$).

We recapitulate.
The condition $Q=0$ at the pole
makes $v$ generally nonzero on the axis.  A non-vanishing $v$
on the rotation axis centrifugally expels matter from the region
$\Theta=0$.  This makes $\et$ zero, leading to a logarithmic
divergence of $P(\Theta)$ at $\Theta=0$ that allows
the automatic satisfaction of $P'=+\infty$ at the north pole,
once we have located it properly relative to the equator.
The latter can be assured by a proper technique to determine
$\Theta_0$ (see below).

Next, we demand the solution to be symmetric about the midplane.
This implies that the first derivatives of metric coefficients
have to vanish there.  Thus, at the midplane,
\begin{equation}
    N' = P' = Q' = 0.\label{3d-boundary:midplane}
\end{equation}
Notice from equation (\ref{log e})
that $\et'$ vanishes automatically once the other three
conditions are met. Equations (\ref{3d-boundary:axis})
and (\ref{3d-boundary:midplane}) are the six boundary conditions
referred to at the beginning of this subsection.

\section{Special Solutions}

\subsection{Newtonian Limit}\label{Newton}

Before we describe the numerical procedure and general solution,
let us pause to discuss the Newtonian limit, where $v$
and $\gamma$ are small compared to unity. This limit should yield
Toomre and Hayashi's result.
In the Newtonian limit, $g_{ij}$ approaches the metric of Euclidean
space in spherical coordinates; $g_{0i}$ vanishes; and $g_{00}
\rightarrow -\exp(2\Phi) \sim -1+O(v^2)$. In our metric, this
procedure translates to
\begin{equation*}
    \gamma, v^2, \et, N = O(n) \ll 1, \qquad
    \Theta = \theta, \qquad \Theta_0 = \frac{\pi}{2}, \qquad P = \ln
    \sin\theta, \qquad Q = Z = 0.
\end{equation*}
With these simplifications, Einstein's field equations reduce
to the usual Poisson's relation, and the equations of motion
recover those of Newtonian fluid dynamics.
Instead of \eqref{log e}, we will revert to using \eqref{energy}
to determine the energy density.  Equations (\ref{force_balance}),
(\ref{energy}), and (\ref{N'}) now have the approximate forms:
\begin{equation}
\begin{split}
    &0 = -n+v^2+2\gamma\\
    &\gamma\et = n^2+ n \cot \theta  N' - \frac{1}{4}{N'}^2,\\
    &N'' = \et - N' \cot \theta - 2n.\\
\end{split}
\end{equation}
These equations may be combined to give a single nonlinear first-order
ODE for $N'$:
\begin{equation}
    N'' = \prn{\frac{n}{\gamma}-1}N' \cot \theta - \frac{1}{4\gamma}
    {N'}^2 + \frac{n^2}{\gamma} - 2n,
\end{equation}
to be solved subject to the boundary condition
$N'=0$ for $\theta=0$.  The
appropriate solution is then
\newcommand{\sech}{\text{\,sech}}
\begin{equation}
\begin{split}
    n & = v^2 + 2\gamma,\\
    \et &= \frac{n^2}{\gamma} \csc^2\theta \sech^2\brk{\frac{n}{2\gamma}
    \ln \cot \frac{\theta}{2}},\\
    N' &= 2 n\left\{ \cot \theta  - \csc \theta
    \tanh \brk{\frac{n}{2\gamma} \ln \cot \frac{\theta}{2}}\right\}.
\end{split}
\end{equation}
The first equation recovers the usual Newtonian simplification
for mechanical equilibrium: gravity $n$
is balanced by centripetal acceleration $v^2$ and the specific
pressure gradient $2\gamma$, with
no relativistic corrections for inertia or frame dragging.
The second and third equations give,
respectively, the density distribution (proportional to $\et$) and the
self-consistent gravitational field (proportional to $N'$)  required to achieve
mechanical equilibrium in the $\theta$ direction.  Apart from notational
differences, these results are indeed what Toomre and Hayashi found.

For the nonrotating configuration,
we expect a spherically symmetric solution.  In
this limit, $n = 2\gamma$, and
\begin{equation*}
    \et = 4\gamma, \qquad N' = 0.
\end{equation*}
Expressed in dimensional variables, this is the familiar result,
$\rho = \varepsilon/c^2 = a^2/2\pi G r^2$, appropriate for a SIS.

When there is even a slight amount of rotation, $n>2\gamma$,
and $\et \rightarrow 0$ on the axis.  In
fact, for $\theta \ll 1$,
\begin{equation*}
    \et \propto \theta ^{n/\gamma -2} \propto \theta^{v^2/\gamma},
\end{equation*}
and curves of constant density $\rho \propto \et(\theta)/r^2$
follow
\begin{equation}
r \propto \theta^{v^2/2\gamma}.
\end{equation}
Thus, except for a very narrow range of angles in the meridional plane
near the pole, where isodensity contours plunge toward
the origin, isodensity contours otherwise
look nearly circular, $r\approx constant$, when $v^2/2\gamma \ll 1$.
As $v^2/2\gamma$ increases because of a greater importance
of rotational compared to pressure support, isodensity contours become
more flattened toward the equatorial plane.
Indeed, the figures in Toomre (1982) and Hayashi et al. (1982)
show that the solutions rapidly approach
a disk-like solution as one increases the level of rotation
to $v^2/2\gamma \gg 1$.  We expect the same qualitative behavior
to extend into the fully relativistic regime.

\subsection{Relativistic SISs}\label{relSIS}

In the absence of rotation, the general relativistic solution
is a spherically symmetric one with $Q=0$ and $v=0$.
Equations (\ref{force_balance}) and (\ref{energy}) then yield
\begin{equation}
    n = \frac{2\gamma}{1+\gamma}, \qquad
    \et = \frac{4\gamma}{(1+3\gamma)^2}. \label{spherical}
\end{equation}
Equation (\ref{log e}) is now trivially satisfied, whereas the right-hand side
of equation (\ref{N'}) is identically zero, implying that $N'$
is a constant.  In fact, $N'$ has to be 0 if we are to
satisfy the boundary conditions (\ref{3d-boundary:axis})
and (\ref{3d-boundary:midplane}).  With $N'$, $Q$, and $v$ all zero,
the equation for $Z'$ shows that it is also 0. Without loss of
generality, we may then choose the constants $N$ and $Z$ themselves to be 0,
so that the metric (\ref{metric}) involves no extra scale factors
in an interpretation of $r$ as the radial distance from the
origin along any path of constant $(t,\theta, \phi)$.  Thus, the
only nontrivial metric coefficient is $P$, which satisfies the
ODE (\ref{P}):
\begin{equation}
P'' + {P'}^2 = -\alpha^2, \label{alphasq}
\end{equation}
where the positive constant $\alpha \le 1$ is defined as 
\begin{equation}
\alpha \equiv \sqrt{ 1-2\et \gamma} = \frac{\sqrt{1+6\gamma+\gamma^2}}
{1+3\gamma}.
\end{equation}
Equation (\ref{alphasq}) may be put into the form
\begin{equation}
\left( e^P\right)'' = -\alpha^2 e^P,
\end{equation}
and may be integrated, subject to the pole boundary condition
$e^P =0$ at $\Theta = 0$, to yield
\begin{equation}
e^P = A\sin(\alpha \Theta),
\end{equation}
with $A$ a nonzero constant.

When we apply the equatorial boundary condition,
$P' = 0$ at $\Theta = \Theta_0$, we obtain the identification
$\cos \alpha \Theta_0 = 0$, or
\begin{equation}
\Theta_0 = \frac{\pi}{2\alpha}.
\end{equation} 
The only remaining task is to determine the value
of the constant $A$.  We accomplish this task by requiring that,
in the presence of spherical symmetry,
half the circumference of a great circle in the meriodional plane,
$2r\theta_0 = 2r\Theta_0/(1+n)$ according to the metric ({\ref{metric}),
must be equal to half the circumference of a great circle in
the equatorial plane, $\pi r e^{P(\Theta_0)} = A\pi r$.
This equality generates the identification,
\begin{equation}
A = \frac{1}{\alpha (1+n)} = \frac{1+\gamma}{\sqrt{1+6\gamma+\gamma^2}}
\le 1.
\end{equation}
Thus, the ratio of the circumference of a great circle $2A\pi r$ to
its radius $r$ from the origin is less than $2\pi$ for
a relativistic SIS with $\gamma > 0$.  This is a familar
result known for relativistic, self-gravitating, pressure-supported
(and therefore spherically symmetric) equilibria,
but it is usually derived for Schwarzschild (or Oppenheimer-Volkoff)
coordinates, where $2\pi r_S$ is {\it defined} as the circumference of a great
circle, and $r_S$ is {\it not} the true radial distance from the origin.
What appears as distortions of angles in our metric (e.g.,
$\theta_0 \ge \pi/2$ is the ``angle'' between pole and
equator) transforms as distortions of the radial dimension
in the Schwarzschild metric.  If we were to adopt the
Schwarzshild description for angles, the radial distance $r$
to the origin would turn out to be a power law of $r_S$, with an
exponent different from unity.

As a check that our definitions of $\theta$ and $\phi$ lead
to no true distortions of angular relationships in the spherically
symmetric case, we note that the surface area of a sphere
can be calculated from the metric (\ref{metric}) as
\begin{equation}
2r^2\int_0^{2\pi}d\phi \int_0^{\theta_0}e^{P}\, d\theta
= \frac{4\pi r^2}{1+n}\int_0^{\Theta_0} A\sin (\alpha \Theta)\, d\Theta
= \frac{4\pi r^2}{\alpha^2(1+n)^2}.
\end{equation}
If we compare this formula to the circumference of a great circle
$2\pi r/\alpha (1+n)$, we note that the ratio of the surface area of
a sphere to the square of the circumference of a great circle on the
surface of that sphere equals $1/\pi$, the same relation as for
Euclidean geometry (QED).  Angles (polar and azimuthal)
maintain their usual relations; it is only the radial direction
that suffers true distortion in a relativistic SIS.  This completes
our discussion of SITs with analytically soluble forms.

\section{Numerical Solution}\label{3d-Numerical}

From dimensional considerations, the solutions for rotating SITs
form a two-parameter family. The equation of state is a result of
microscopic physics, which is independent of the macroscopic
spacetime geometry.  Thus, the square of the isothermal sound
speed, $\gamma$, is a natural choice for one of the parameters.
The other convenient constant of the problem, $n$, measures the
strength of the gravitational field.  Ideally, we might have
preferred to specify in advance the amount of rotation.  However,
this is inconvenient as $v$ is a function of ``polar angle'' $\Theta$.
We therefore fix on $\gamma$ and $n$ to parameterize our
solutions, and let this combination determine implicitly the
amount of rotation on the equator and elsewhere.

To implement a practical numerical scheme, we solve
equation \eqref{force_balance} as a quadratic relation for $v$ to obtain
\begin{equation}
    v = \frac{-Q(1+\gamma)(1-n) + \sqrt{(1+\gamma)^2 (1-n)^2
    Q^2 + 4(1-\gamma)(n+n\gamma -
    2\gamma)}}{2(1-\gamma)}.\label{v}
\end{equation}
Using the above relation, we compute $v(\Theta)$ from $Q(\Theta)$
at every step for use in equations (\ref{dynamic}).
Causality demands $v<1$ and $\gamma
<1$, and $Q$ must have the same sign as $v$, which is taken to be positive
by convention.  Thus, the solution space is constrained to satisfy
\begin{equation}
    0 \le \gamma \le 1, \qquad \frac{2\gamma}{1+\gamma} \le n \le 1.
    \label{sol space}
\end{equation}
In fact, $(1+\gamma)/(1-\gamma)$ times $n-2\gamma/(1+\gamma)$
is the value of $v^2$ at the pole.

For convenience, we start at the midplane.
Because the governing ODEs,
equations (\ref{dynamic}) and (\ref{constraint}), contain
no coefficients that depend explicitly on $\Theta$ itself, these equations
are formally invariant to an overall translation of the value
of $\Theta$ at the midplane.
Thus, a simple transformation of variables
from $\Theta$ to $\Psi \equiv \Theta-\Theta_0$ means that
we do not need to guess a value of $\Theta_0$ at the equator,
but can obtain it afterwards when we reach the pole, as defined by
the value of $-\Psi$ where $Q=0$ is reached.  Attainment of
$Q=0$ at the pole leads to the automatic satisfaction there of
$P'=+\infty$ as we discussed earlier.

The boundary conditions \eqref{3d-boundary:midplane} define the corresponding
values of $N^\prime$, $P^\prime$, and $Q^\prime$ at the new
midplane label $\Psi=0$.
In principle, we need to guess midplane values of
$Q$ and $\et$ to begin the integration toward the north pole.
In practice, we can eliminate the need to guess $\et$ at the equator, and use
equation (\ref{energy}) to obtain this value, given $\gamma$
and $n$.   Thus, we are left
with a one-dimensional shooting task of
adjusting a guessed midplane value of $Q=q$ to satisfy
the boundary condition $N'=0$ at the pole $\Psi=-\Theta_0$.
Since we want the local energy density to
be positive in equation (\ref{energy}), a useful
value of $q$ must reside within the interval:
\begin{equation}
  0 \le  q \le \frac{2n}{1-n}. \label{positive_energy}
\end{equation}

A slight difficulty enters to complicate the above program
that deserves mention.  When regarded as a function of $q$,
the relevant zero of $N^\prime(0)$ turns out to be a double root; i.e.,
$\partial N^{\prime}(\Theta=0)/\partial q$ = 0 at the
sought-after value of $q$.  We therefore numerically compute
$\partial N'(\Theta=0)/\partial q$ and refine the grid of
possible $q$ values when this quantity becomes small, so as
not inadvertently to jump over the root of $N'(\Theta=0)$.

\section{Results}\label{3d-Results}

To make meridional cross-sectional plots of isodensity surfaces,
we use ``cylindrical'' coordinates $(t, \varpi, \phi, z)$ that
are the counterparts of the
system used by Lynden-Bell \& Pineault (1978a,b):
\begin{equation}
    \varpi = r^{1/k}\sin (\theta/k), \qquad
    z = r^{1/k}\cos (\theta/k), \label{LB-P}
\end{equation}
where
\begin{equation}
    k \equiv \frac{2}{\pi}\theta_0
\end{equation}
is the factor $\ge 1$ for the models of this paper
discussed in \S \ref{geometry}.

In these coordinates the metric reads
\begin{equation}
\begin{split}
        ds^2 &= -(\varpi^2+z^2)^{kn} e^N dt^2 +
    (\varpi^2+z^2)^{k} e^{2P-N} [d\phi - (\varpi^2+z^2)^{k(n-1)/2} e^{N-P}
    Q dt]^2 \\
&+k^2(\varpi^2+z^2)^{k-1}e^{Z-N} (d\varpi^2 +dz^2),
     \label{cylinmetric}
\end{split}
\end{equation}
with the $\theta$ dependences of the metric coefficients
$N(\theta)$, $P(\theta)$, $Q(\theta)$, $Z(\theta)$ now to be
interpreted by making the substitution
\begin{equation}
\theta = k\arctan (\varpi/z).
\end{equation}

Because of spatial curvature, {\it faithful}
representations of both lengths and angles (in the meridional
plane) would require curved sheets of paper or a 3-D embedding diagram.
Conventional publishing
limitations restrict our ability to do full justice to
such possible graphical representations, and we have compromised
by making $(\varpi,z)$ plots of the isodensity contours
using the coordinate definitions (\ref{LB-P}).  We emphasize,
however, that when $n$ is not small (i.e., when gravity is not
weak), such plots give graphical relationships of isodensity contours
for the {\it coordinate labels only},
and do not accurately represent spatial separations
or angles.  Figure \ref{isopycnic} gives such plots for
the case $\gamma = 1/3$ (e.g., an ultrarelativistic
{\it ideal} gas whether ``isothermal'' or not) when $n$ ranges from a minimum
value $n=1/2$, corresponding to no rotation (i.e., a SIS),
to a value close to the maximum value $n=1$, corresponding
to the most rapidly rotating SIT possible.

When we introduce rotation, $\et$ has to vanish on the axis as a
result of the centrifugal emptying process described earlier.
Near the pole, we assume
$\et \propto \Theta^\lambda$ for $\lambda >0$, and adopt a series
expansion for $P'$, using the ``dynamical'' equations
\eqref{dynamic}:
\begin{equation*}
    P' = \frac{1}{\Theta} + O(\Theta), \qquad \lambda =
    \left( \frac{v^2}{\gamma}\right)\left(\frac{1+\gamma}{1-v^2}\right)
    = \frac{n+n\gamma
    - 2\gamma}{\gamma(1-n)} = \frac{1+\gamma}{\gamma(1-n)}(n-n_{\text{min}}),
\end{equation*}
where we have used equation \eqref{force_balance} on the axis.
Notice that in our constrained solution space \eqref{sol space},
$\lambda \ge 0$.  For a slight rotation, $\lambda \ll 1$, the
power-law behavior of the
energy density near the axis causes it to change at a normalized
radius from $1$ to $0$ nearly
discontinuously as the pole is approached. This behavior is
depicted in Figure \ref{energy-angle}. Expressed differently,
the parameter $\lambda$ controls the slope
of isodensity contours near the pole.
As $\lambda \rightarrow 0^+$, the isodensity curve in the meridional
plane switches from being horizontal to being vertical very quickly
as the curve approaches the origin (see Fig. \ref{isopycnic}).
Although $\et$ has to vanish formally on the
axis, Figure \ref{isopycnic} shows isodensity contours essentially
to be spherical when $\lambda$ is small.

In general, for positive values of $\gamma < 1$,
$\lambda$ ranges from $0$ for the minimum allowed
gravitational index $n$ (and no rotation) to $\infty$ for $n=1$,
which is the maximum gravitational field strength allowed by
causality.  When $\gamma = 1$, a limiting process is required.  If
we take the limit $\gamma \rightarrow 1$ first, then we are forced
to take $n=1$, and we recover the nonrotating solution of a
self-similar sphere with $\lambda = 0$. However, if we fix $n=1$
and let $\gamma \rightarrow 1$, then the result is a maximally
rotating SIT with $\lambda = \infty$.  This ambiguity
is a manifestation of the sphere-toroid transition at zero
rotational velocity discussed earlier.

From a purely mathematical point of view, the double-limit
ambiguity arises because of the non-uniform convergence of
solutions towards the singular point $n=\gamma=1$.  After all, the
Einstein equations formally allow solutions with $n>1$.  However,
these spacetimes are acausal and have divergent energy density on
the axis, and thus are irrelevant for physical considerations. We
shall not digress further on this issue.  In the following
analysis, it is understood that we always take $\gamma\rightarrow
1$ first.

\subsection{Velocity Field}

With the three-dimensional solutions, we have a chance to study
the velocity field as a function of polar angle.  In a disk
of infinitesimal thickness, for each value of $\gamma$,
the rotation speed in the plane of the disk is limited to
a maximum value that arises when the frame dragging becomes
infinite: $Q \rightarrow \infty$ accompanied by $n\rightarrow 1$.
Empirically, CS (see their Table 1) found
the maximum rotational velocity for a disk to be
a function of $\gamma$ given approximately by
\begin{equation}
    v_c = \frac{1}{2.294+1.091\gamma}. \label{vcrit-empirical}
\end{equation}
The fact that the maximum velocity \textit{any} disk can have is
only about $43.8\%$ of the speed of light came as a surprise when
Lynden-Bell \& Pineault (1978b) first discovered it for
a cold disk, and it has remained a mystery until now.

Three-dimensional SITs are more realistic systems than
completely flattened SIDs (with their infinitesimal thickness
resulting from an artificial imposition of a highly anisotropic
pressure tensor). In a SIT, as seen in Fig \ref{velocity}, the
velocity field increases towards the axis, reaching a significant
fraction of the speed of light in most cases. This
qualitative result was anticipated
by our previous discussion.  The maximum of
the velocity field occurs on the axis where $Q=0$ and is given by
equation (\ref{v}) as
\begin{equation*}
    v_{\text{axis}}^2 = \frac{n+n\gamma-2\gamma}{1-\gamma}.
\end{equation*}
For a fixed value of $\gamma\in [0,1)$, $v_{\text{axis}}$
increases from $0$ for minimum rotation to $1$ (the speed of
light) for the ultrarelativistic limit.  The sequence is then
terminated while the equatorial velocities are still a minor
fraction of the speed of light, comparable, in fact to the
subluminal values given by equation \eqref{vcrit-empirical} for
the pure-disk solutions.  (For more precise comparisons, see
below.)  The naive expectation that relativity sets limitations on
physical systems only because matter acquires velocities
approaching the speed of light is indeed met for SITs where models
terminate when (vanishingly small amounts of) gas near the pole
become ultra-relativistic.  It is only SIDs that present an
apparent puzzle because the assumption of a vacuum above and below
the exact midplane of the disk removed all material tracers of
this ultra-relativistic motion a priori!

The restriction of equatorial rotation velocities to
modest fractions of the speed of light implies that
differentially rotating, gaseous configurations
with isotropic pressure contributing a significant
fraction of the total support against self-gravity,
cannot be too highly flattened.  Despite this caveat,
however, it is remarkable how useful the approximation
of a disk-like geometry can be to predict certain
physical characteristics of more realistic systems.
In Table 1, we tabulate as a function of $\gamma$
the maximum equatorial velocity $v_c$ attainable by
a SID or SIT when the gravitational field acquires the
maximum index $n=1$ (reached by extrapolation from values
of $n$ somewhat smaller than unity).  From this tabulation, we
see that the two values of $v_c$ are in very good agreement
when $\gamma \ll 1$.  This is a natural result because the critically
rotating SIT is extremely flattened at low levels of
pressure support.  However, the good agreement persists
to relativistic values of $\gamma \approx 0.5$, a
surprise since $v_c^2/\gamma$ is far from being large
compared to unity, as is required formally for the disk
approximation to hold.  Indeed, complete breakdown of
the disk approximation does not occur in the Table until
$v_c^2/\gamma$ is significantly {\it smaller} than unity.
Recalling that the ratio of pressure to energy density (including
rest mass) is given by $\gamma = 1/3$ for ultrarelativistic ideal gases,
we see that Table 1 implies excellent agreement in the
gross properties of SITs and SIDs with realistic equations
of state which are maximally rotating.
(The apparent discrepancy at $\gamma = 0$ arises from an imperfect
extrapolation of SITs to this limit.)
This finding relieves some of the worries that one might otherwise
have about the applicability of results derived from a flat
disk analysis when parameters are used that would
give such disks appreciable vertical thickness if
the pressure were isotropic rather than anisotropic
(see the caveats expressed in CS).

\section{Conclusion and Commentary}\label{conclusion}

With the aid of self-similarity, we have constructed a two
parameter family of semi-analytic solutions to the Einstein field
equations, parameterized by the gravitational index and an
isotropic pressure.  The isodensity surfaces are toroids,
qualitatively similar to their Newtonian counterparts
(Toomre 1982, Hayashi et al. 1982).

The main difference between relativistic SITs and their non-relativistic
counterparts is that the former do not satisfy, even approximately,
the Poincar\'e-Wavr\'e theorem (see, e.g., Tassoul 1978).
In the Newtonian limit, Goldreich \& Schubert (1967) have shown that
equilibria which violate the constraint of isorotation on cylinders
are unstable on timescales that range from dynamical to secular.
It is not known whether such instabilities persist into the
relativistic regime if the cause of the departures from isorotation
on cylinders is the dragging of inertial frames.  This issue
deserves further investigation.

We have also gained insight into how frame dragging figures
prominently into another puzzle.  Lynden-Bell \& Pineault (1978b)
found, and CS confirmed, that the rotational velocities of completely flat,
relativistic disks are limited to be less
than $0.438$ times the speed of light.
This limitation appears quite peculiar, but when we examine the behavior of
three-dimensional toroids, the velocity field turns out, as summarized
above, to be a function of the
polar angle.  While the equator is still rotating modestly (subluminally),
the velocity on the axis can reach the speed of light when an appropriate
combination of parameters holds.  As a result, matter is pushed
away from the axis much more strongly than one might have suspected
by examining only the ratio of rotational support $v^2$ on the
midplane to the pressure contribution $\gamma$.  As a byproduct, a disklike
approximation still holds for many SITs to a surprisingly good degree
when one might otherwise have expected a complete breakdown of such a
simplified description.

The direct applicability of relativistic SITs and SIDs to
realistic astrophysical systems is questionable. They have inner
and outer singularities (infinite central density, divergent
enclosed mass at infinity) that lead to spacetime geometries
(quasi-naked singularity at the origin, nonflat spacetime
asymptotically) which are conventionally rejected as inapt
descriptions of reality. However, we note that the issue of the
possibility of naked singularities in the universe has not been
settled in any definitive form (Penrose 1998), and non-asymptotically flat
spacetimes are a common feature of many standard cosmologies,
including the currently favored versions of accelerating universes
with positive cosmological constants. Moreover, for a problem to
be interesting in general relativity, spacetime has to possess
curvature.  But self-similar spacetimes that have power law
dependences on some radial coordinate are guaranteed to run into
trouble at small and large radii, since the interesting curvature
in the body of the problem has to be carried into the origin and
to infinity (where they become ``objectionable'').

Our point here is not to argue for
the direct application of the self-similar models of
the current line of research to any known astrophysical
object, but to point out their possible utility in the study of
a restricted but extremely interesting subset of problems
in general relativity: the role of frame dragging, nontrivial
equilibria with and without convenient symmetries, gravitational
collapse, the formation of singularities, the efficiency of the
generation of gravitational radiation, etc.  Computational results
in this field are hard to obtain as it is; we should
not shrink from applying a technique -- self-similarity -- that has
proven itself very fruitful in application to many subjects other than general
relativity, simply because its assumption produces a few
blemishes in otherwise acceptable physical models.

\acknowledgements

This work has been supported in the United States by the National Science
Foundation by a Graduate Fellowship awarded to MJC and by
grant AST-9618491 awarded to FHS.  In Taiwan, we are pleased to
acknowledge the support of the National Science Council.

\newpage
\begin{figure}[ht]
\begin{center}
\mbox{\subfigure{\includegraphics[width =
.45\textwidth]{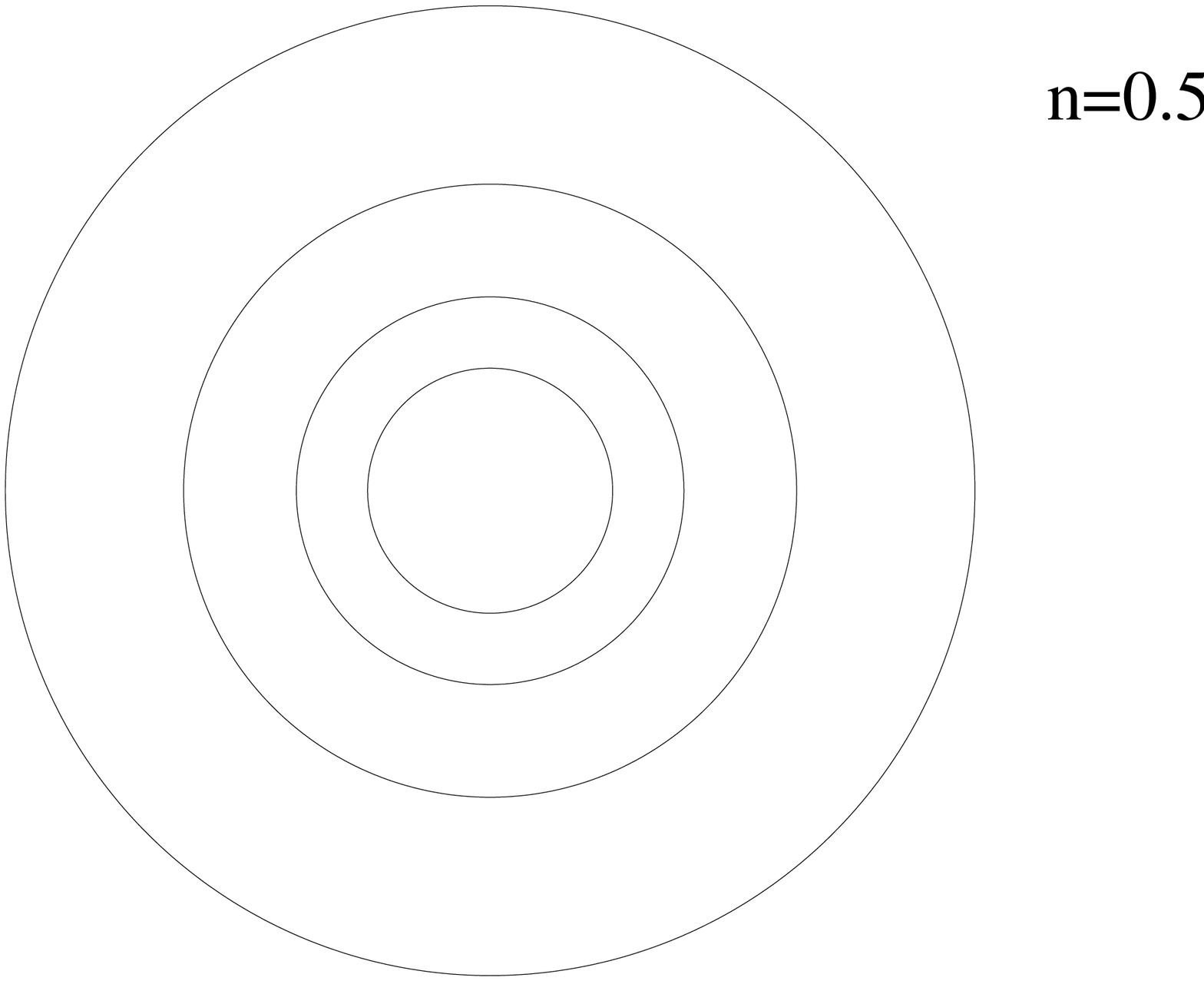}}}
\mbox{\subfigure{\includegraphics[width =
.45\textwidth]{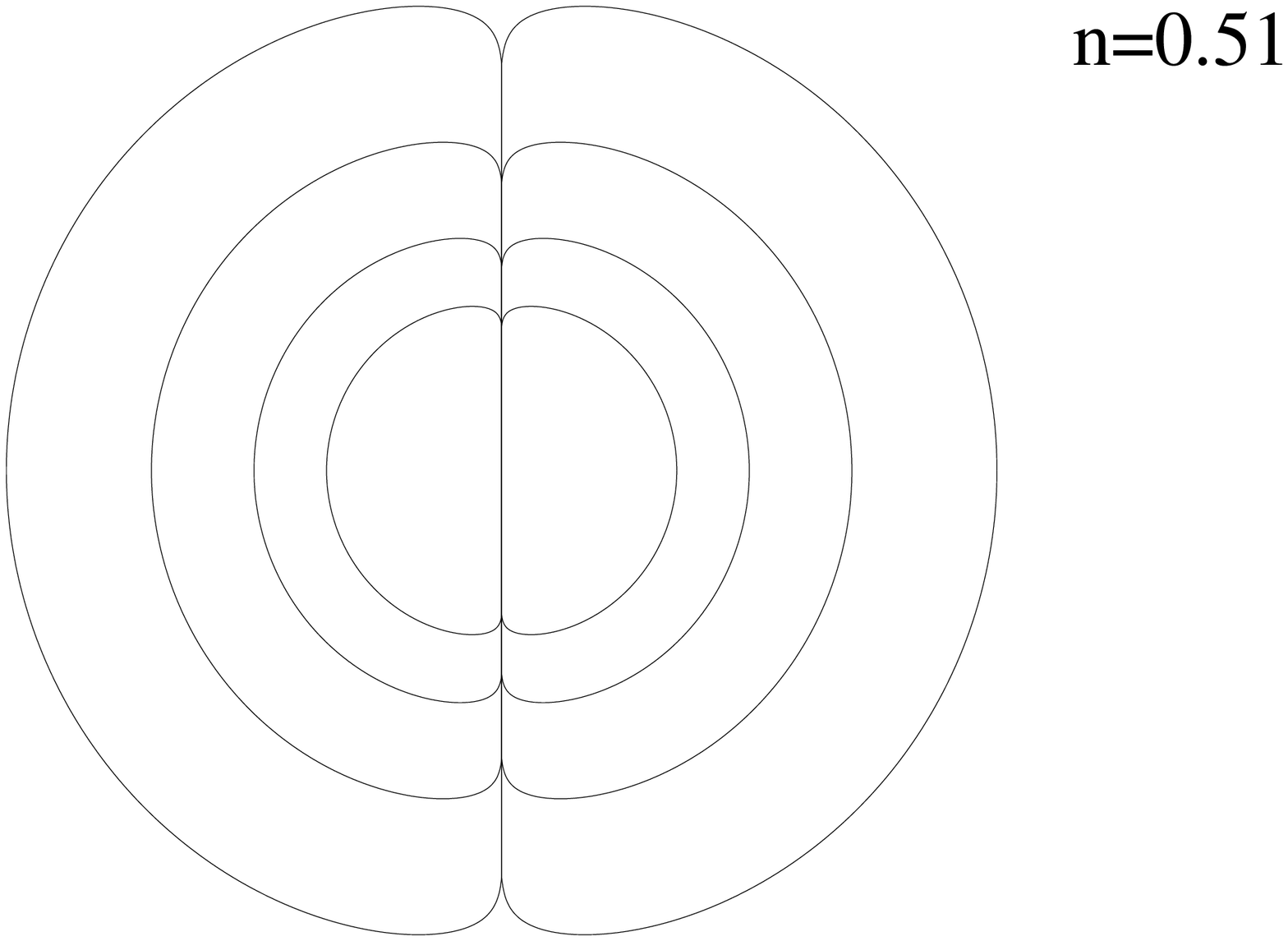}}}
\mbox{\subfigure{\includegraphics[width =
.45\textwidth]{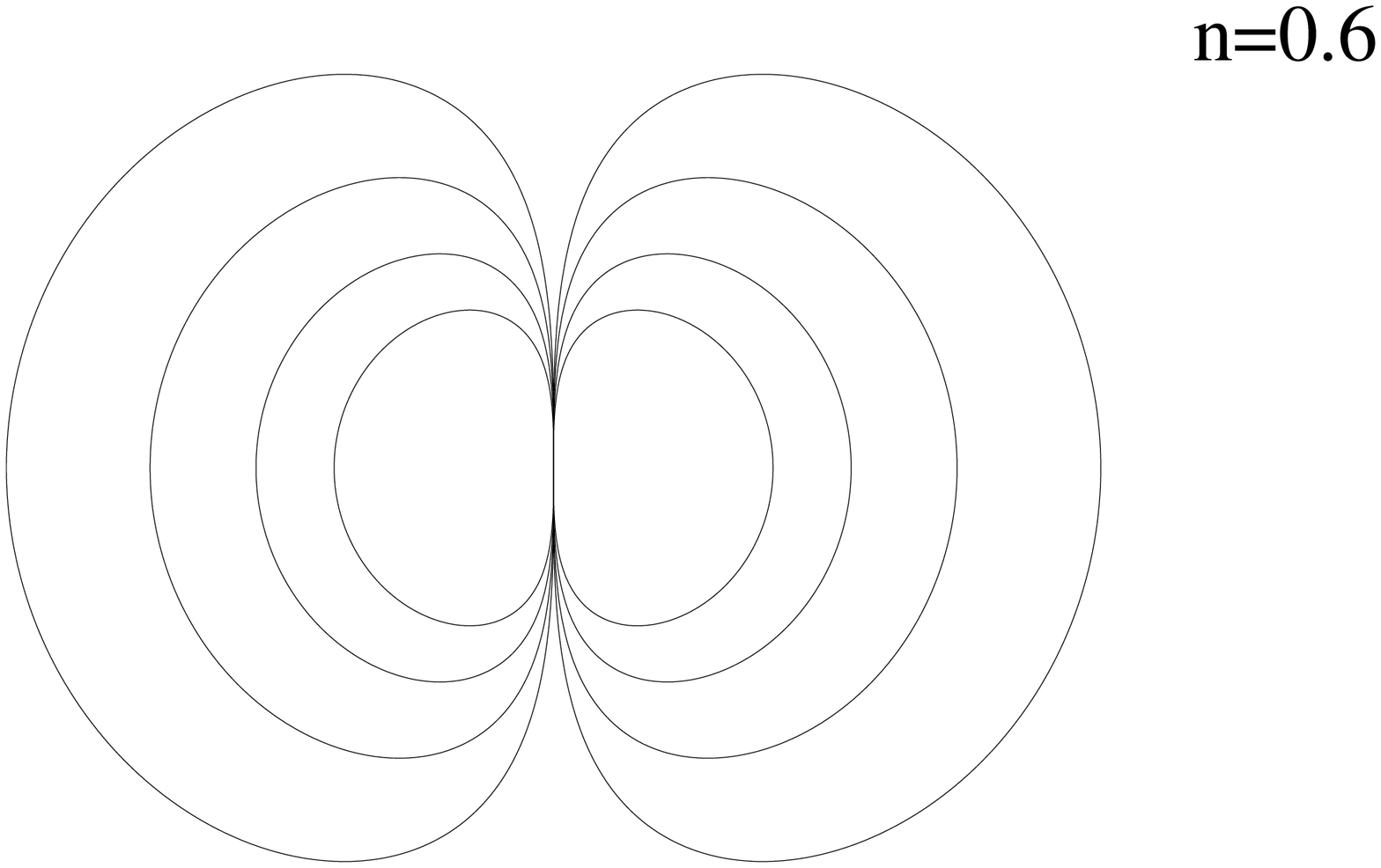}}}
\mbox{\subfigure{\includegraphics[width =
.45\textwidth]{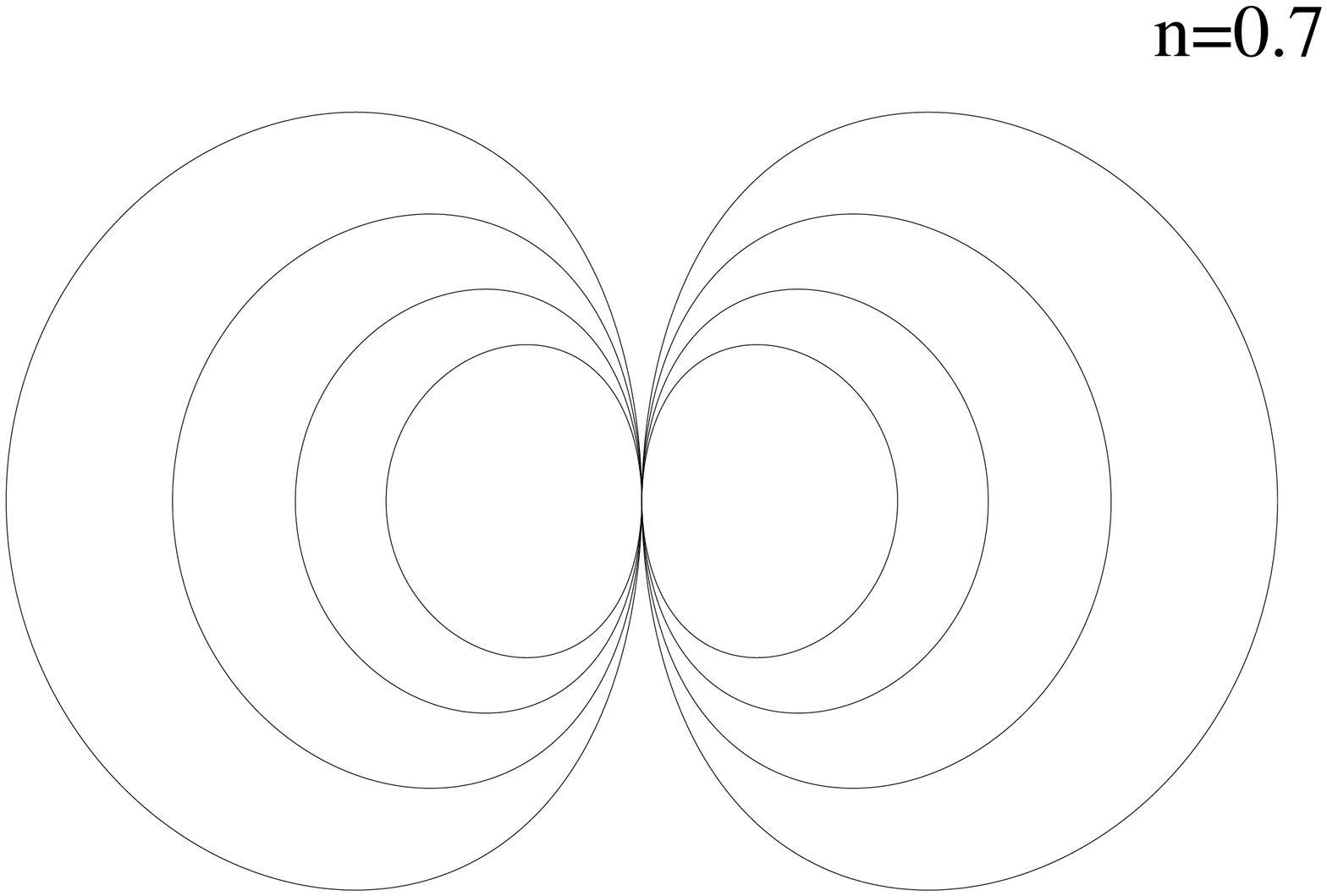}}}
\mbox{\subfigure{\includegraphics[width =
.45\textwidth]{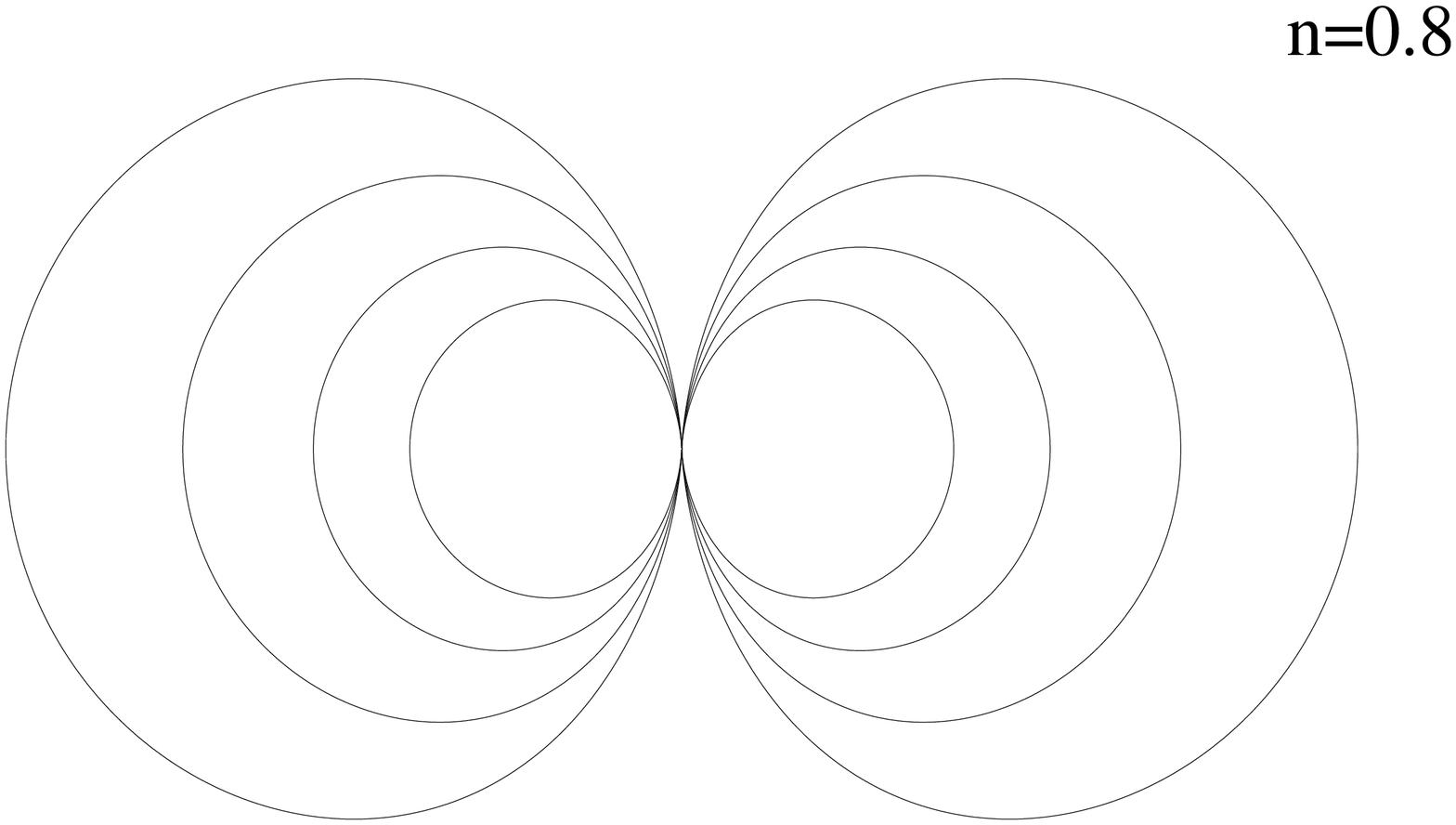}}}
\mbox{\subfigure{\includegraphics[width =
.45\textwidth]{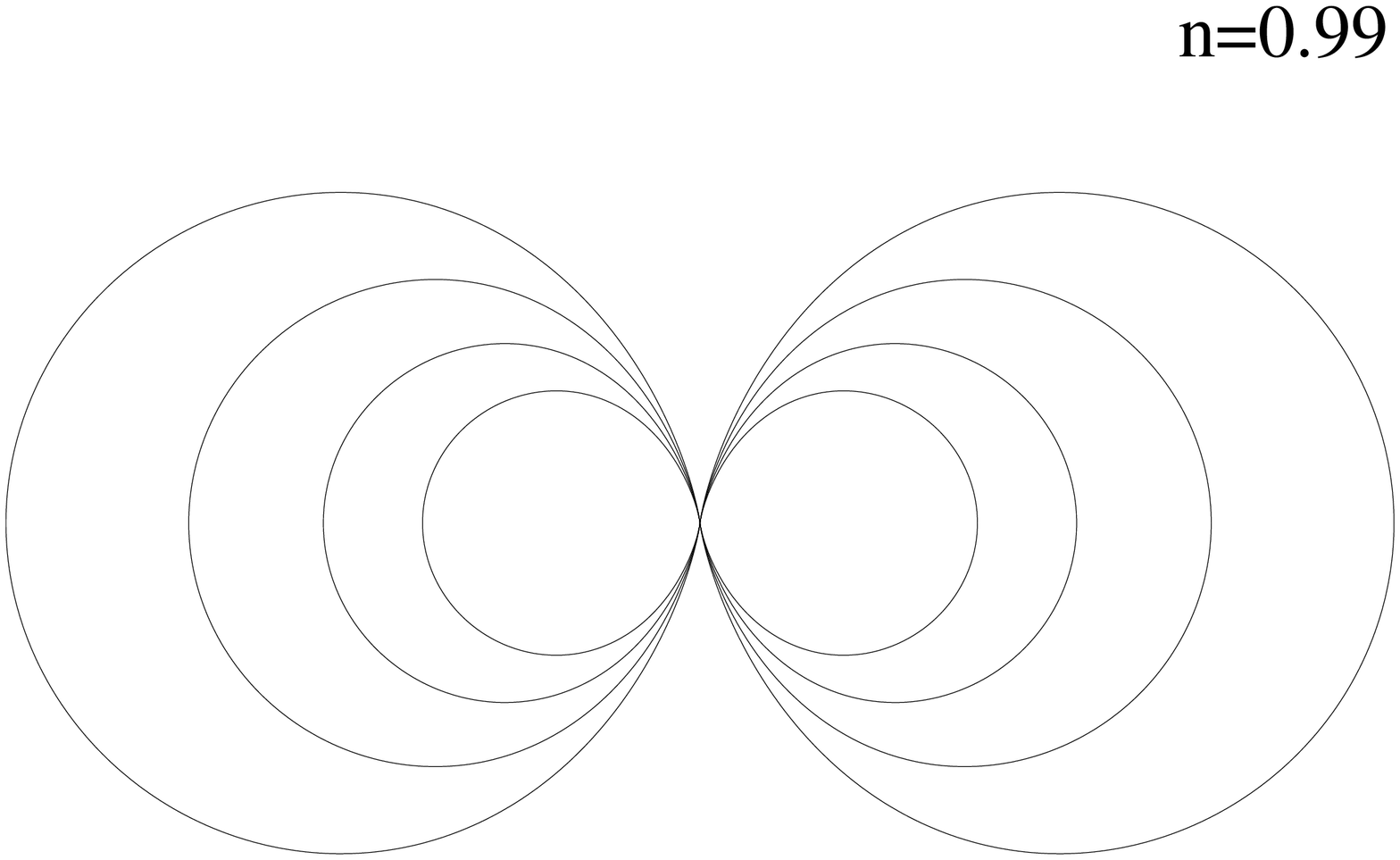}}}
\end{center}
\caption{Meridional cross section of constant energy surfaces for
$\gamma=1/3$.  The contours corresponds to $1$, $2$, $4$, and $8$
times some reference value. The allowed values of gravitational index
range from $n=2\gamma/(1+\gamma)=0.5$ to $n=1$.}
\label{isopycnic}
\end{figure}

\begin{figure}[ht]
\begin{center}
\includegraphics[height = .8\textwidth, angle = 90]{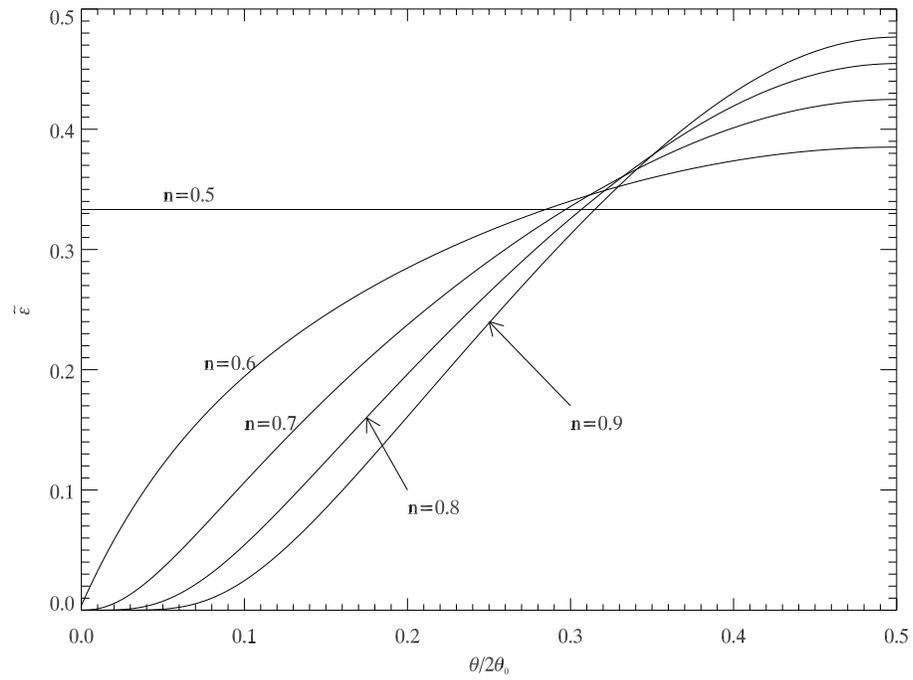}
\end{center}
\caption{Rescaled energy density $\et$ as a function of polar
angle for $\gamma = 1/3$.  The minimum gravitational index $n=0.5$
corresponds to a nonrotating sphere with constant $\et$.  As $n$
increases, energy is redistributed toward the midplane.}
\label{energy-angle}
\end{figure}

\begin{figure}[ht]
\begin{center}
\includegraphics[height = .8\textwidth, angle = 90]{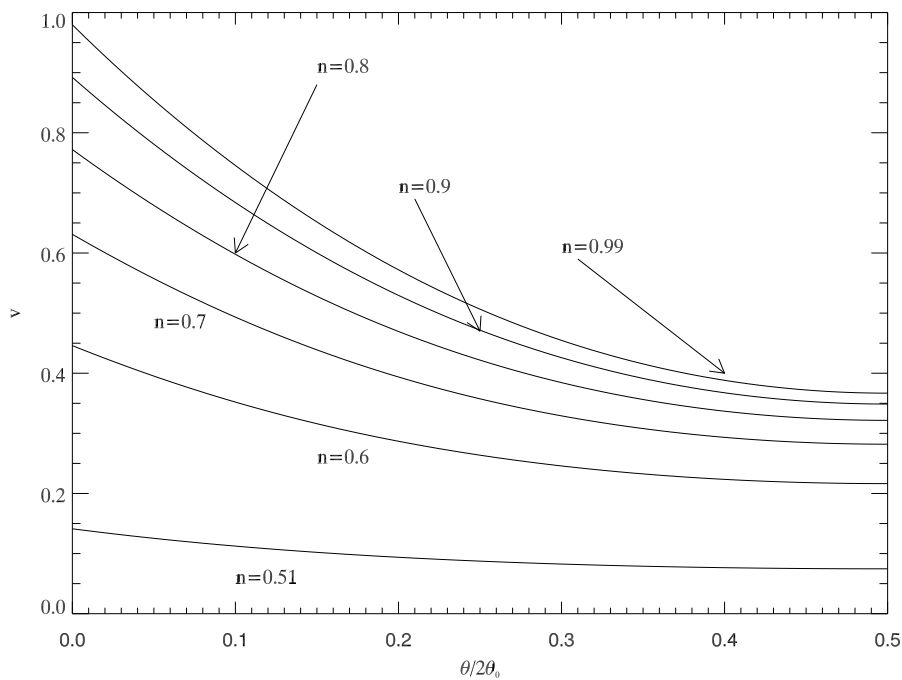}
\end{center}
\caption{Linear rotational velocity as a function of polar angle
for $\gamma = 1/3$.} \label{velocity}
\end{figure}

\begin{deluxetable}{ccc}
\tablecaption{Critical Velocities for SID and SIT as a Function of
Sound Speed Squared} \tablecolumns{3} \tablewidth{0pt}
\tablehead{\colhead{$\gamma$}&\colhead{SID $v_c$}&\colhead{SIT
$v_c$}}
\startdata
        0.0&0.438&0.439\cr
        0.1&0.415&0.414\cr
        0.2&0.398&0.395\cr
        0.3&0.381&0.375\cr
        0.4&0.366&0.347\cr
        0.5&0.351&0.313\cr
        0.6&0.339&0.276\cr
        0.7&0.327&0.235\cr
        0.8&0.316&0.190\cr
        0.9&0.306&0.132\cr
        1.0&0.298&0.050\cr
\enddata
\end{deluxetable}
\end{document}